\documentclass[reprint,onecolumn,amsmath,amssymb,aps,
]{revtex4}

\usepackage{epsfig,bm,epsf,graphics}
\usepackage{graphicx}
\usepackage{dcolumn}
\usepackage{bm}
\usepackage{xcolor}

\begin{document}
\preprint{APS/123-QED}

\title{Magnetic properties of perovskites Pr$_{0.9}$Sr$_{0.1}$Mn$_{0.9}^{3+}$Mn$_{0.1}^{4+}$O$_{3}$: Monte Carlo simulations and experiments}

\author{Yethreb Essouda$^{1}$\footnote{yethreb.essouda@gmail.com},  Hung T. Diep$^{2}$\footnote{diep@cyu.fr, corresponding author},  Mohamed Ellouze$^{1}$\footnote{mohamed.ellouze@fss.usf.tn}, and E. K. Hlil $^3$\footnote{el-kebir.hlil@neel.cnrs.fr}}
\affiliation{%
$^1$Sfax University, Faculty of Sciences of Sfax, LM2EM, B.P. 1171, 3000, Sfax, Tunisia.\\
$^2$  Laboratoire de Physique Th\'eorique et Mod\'elisation,
CY Cergy Paris Universit\'e, CNRS, UMR 8089\\
2, Avenue Adolphe Chauvin, 95302 Cergy-Pontoise Cedex, France.\\
$^3$ Universit\'e Grenoble Alpes, CNRS, Grenoble INP, Institut N\'eel, 38000, Grenoble, France.\\
}



\date{\today}


\begin{abstract}

This work presents the remarkable experimental magnetocaloric properties  of  the perovskites  Pr$_{0.9}$Sr$_{0.1}$Mn$_{0.9}^{3+}$Mn$_{0.1}^{4+}$O$_{3}$, including the magnetic entropy change $|\Delta S_m|$ and the Relative Cooling Power (RCP). To understand  these striking properties,
we elaborate in this paper a model and use   Monte Carlo (MC) simulations to study it for comparison.  For the model, we take into account nearest-neighbor (NN) interactions between magnetic ions Mn$^{3+}$($S=2$) and Mn$^{4+}$($S=3/2$) and the interactions between these Mn ions with the magnetic Pr  ions.  The crystal is a body-centered tetragonal lattice where the corner sites are occupied by Mn ions and the center sites by Pr and Sr ions in their respective concentrations given in the compound formula.  We use an Ising-like spin model describing a strong anisotropy on the $z$ axis.  We show that pairwise interactions between ions cannot reproduce the large plateau of  the magnetization experimentally observed below the phase-transition temperature. By introducing for the first time a many-spin interaction between Mn ions, we obtain an excellent agreement with experiments. 
Fitting the experimental Curie temperature $T_C$ with the MC transition temperature, we estimate the value of the effective exchange interaction in the system. From this value, we estimate various exchange interactions between ions: the dominant one is that between Mn$^{3+}$ and Mn$^{4+}$ which is at the origin of the ferromagnetic ordering below $T_C$.  We also studied the applied-field effect on the magnetization in the region below and above $T_C$. The obtained MC results for $|\Delta S_m|$ are in agreement with experiments performed  for applied fields from 1 to 5 Tesla. MC results of RCP are also shown and compared to experimental ones. Various other physical quantities obtained from MC simulations including internal energy and  specific heat, versus temperature are also shown and discussed.  
\vspace{0.5cm}
\begin{description}
\item[PACS numbers: 5.10.Ln;64.30.-t;75.50.Cc]
\end{description}
\end{abstract}

\maketitle


\section{INTRODUCTION}

Changes observed in Earth’s climate since the industrial revolution was mostly driven by increase dramatically of global energy consumption for cooling. As well as, using conventional gas compression refrigeration systems have shown dangerous effects on the earth because it is consumed 15\% of electricity generated globally and account for 10\% of global greenhouse gas (GHG) emissions into the atmosphere resulting in a global warming that we have observed over the last century period and which has become an important environmental issue worldwide. 
In 2016, for example, scientists concluded that the global electricity usage for cooling was 2000 TW h, or 18.5 \% of annual electricity consumption [1] and thanks largely demand for cooling, particularly in the world's warmer regions, global energy consumption for cooling could explode tenfold by 2050 [2], which has also motivated the exploration of Photovoltaics (PV) production for cooling [3] to conserve energy and electricity, and alternative coolers to reduce the negative impacts of global warming. Already today, debating global warming and energy efficiency scenarios and input them into globally validated models for cooling demand, photovoltaic (PV) electricity generation, thermal storage and looking for alternative coolers are the most critical blind spots in today's energy debate.
Nowadays, among the most prominent worldwide in the new generation of refrigerants, Magnetic Refrigeration (MR), based on the magnetocaloric effect (MCE), becomes a promising technology and alternative refrigeration systems that can replace the conventional gas refrigeration technique [4]. In contrast to a compression cycle the MR can be an eco-friendly refrigeration technology [5], due to the use of magnetic materials that acts as a refrigerant, which produces no ozone-depleting chemicals (CFCs), hazardous chemicals (NH3), or greenhouse gases (HCFCs and HFCs) [6]. MR has presented another advantage like large refrigeration efficiency and the potential to reduce energy use by 30\% and requires no refrigerant.
The MCE is an intrinsic property of magnetic materials, is defined as the changes in the magnetic entropy ($|\Delta S_m |$) and hence the temperature of the material when placed in a magnetic field under adiabatic conditions [7-9]. MCE was first discovered by Warburg (1881) with iron [10],  it paved the way for further developments. The reader is referred to Ref.  [11] for a very good review of historical developments, both experimentallly and theoretically, during the following 50 years.  Though Gd-based alloys are the most studied materials which exhibits excellent magnetocaloric properties, but its high cost which could be as high as  4,000 kg–1 limits their use as magnetic refrigerant [12]. In fact, a huge part of current researchers is orienting its attention on some new MCE materials having the best magnetic properties at a low production cost. The most promising are the rare-earth transition-metal manganites of general formula Ln$_{1-x}$A$_x$MnO$_3$ (Ln = trivalent rare-earth, A = divalent alkaline earth) because of their exciting properties such as colossal magnetoresistance (CMR) and colossal magnetocaloric effect (MCE) [13-19]. These rare earth manganese oxides also show several other interesting physical properties like magnetodielectric behavior [20], multiferroicity [21], large magneto-optic responses [22]and orbital ordering [23].  We note that new researches [24,25] have been recently carried out to show the high potential of nanopowder perovskites for magnetic refrigeration application, which is urgently crucial for green and environmentally friendly technology.
To get a clear idea about the performance of materials used in magnetic refrigeration devices, the main requirements for a magnetic material are the large magnetic entropy change $|\Delta S_m|$, the large spontaneous magnetization as well as the sharp drop in the magnetization associated with the ferromagnetic to paramagnetic transition at the Curie temperature $T_C$ [26, 27]. As known, large magnetic entropy changes were reported in other materials such as Pr$_{0.6}$Sr$_{0.4}$MnO$_3$ [28]. It is distinguished by a maximum change of magnetic entropy $|\Delta S_m^{max}|$ corresponding to 3.58 J/kg K under a magnetic field of 5T and a Relative Cooling Power corresponding to 159.37 J/kg which make these materials good candidates for magnetic refrigeration applications. Furthermore, many experimental works were made for the perovskite Pr$_{1-x}$Sr$_x$MnO$_3$ material. However, these experimental works allow only a partial understanding of the compound. A better understanding of the roles of each microscopic interaction in of this magnetic system requires more theoretical and numerical studies.
The aim of our work is to study the magnetic and magnetocaloric properties of the perovskite  Pr$_{0.9}$Sr$_{0.1}$MnO$_3$  material by performing experiments and by Monte Carlo simulations to compare to experimantal data. For the modeling, we use an Ising-like spin model with various interactions based on experiment observations. To fit with experiments,  in addition to pairwise interactions between magnetic ions, we introduce a new many-spin interaction term between Mn ions. As seen in this paper, this model is justified by a comparation with the experimental measurements performed on this material.

Before introducing our perovskte compound, let us emphasize that rich and various perovskite compounds due to doping 
are currently attracting a considerable attention [29]. They have many remarkable properties  because of the complex interplay among spins which induces a rich phase diagram [30-33].  There are many applications in particular in spintronics [33-35].
The diagram in the phase space defined by concentration $x$, temperature $T$, magnetic field $H$, superexchange (SE) $J$ is not quite clear yet for different compounds. Jonker and Van Santen [36] have studied ferromagnetic compounds of manganese with perovskite structure. Their  properties can be understood as the result of a strong ferromagnetic (positive) exchange interaction
between nearest neighboring Mn ions via intercalated oxygen: Mn$^{3+}$-O-Mn$^{4+}$.\\
Note that the double exchange (DE) mechanism developed by Zener [37-39] explains the existence of ferromagnetism and the metallic behavior at low temperatures. There is now a consensus to recognize that the
interesting properties observed in perovskites are fundamentally originated from the DE mechanism along the link Mn$^{3+}$-O-Mn$^{4+}$. This characteristics is at the origin of a new interesting observed phase
transition in doped manganites [40,41] from a magnetically-ordered phase to the disordered phase. Recent refinement of experimental techniques and the
improvement of the sample quality have made possible to discuss critical phenomena of this transition [42].

\indent In this paper we experimentally investigate magnetic properties of perovskite manganite Pr$_{0.9}$Sr$_{0.1}$Mn$_{0.9}^{3+}$Mn$_{0.1}^{4+}$O$_{3}$ using by Powder X-ray diffraction (XRD) technique and the SQUID magnetometer developed in Louis Neel Laboratory at Grenoble, France. In order to fit with the experimental results, we propose a new model including a many-spin interaction in addition to the pairwise exchange interactions between magnetic ions. We then study the model  with Monte Carlo (MC) simulations.
Note that the compound Pr$_{0.55}$Sr$_{0.45}$MnO$_{3}$ has been experimentally studied by Fan et al [43] and by Saw et al [44]. Other concentrations of Pr, namely Pr$_{0.5}$Sr$_{0.5}$MnO$_{3}$  and Pr$_{0.6}$Sr$_{0.4}$MnO$_{3}$, have been studied in Refs. [45,46,47].

We use a discrete spin model to express the strong Ising-like anisotropy along the $z$ axis and we take into account various types of interactions between spins of Mn$^{4+}$, Mn$^{3+}$ and Pr$^{3+}$.  This model is justified by an excellent agreement with experimental magnetization measurements performed on this material.\\

\indent The paper is organized as follows: in section \ref{experiment}, we describe our experiment.  Our theoretical model and the MC method are shown in section \ref{model}.  Results are shown and compared to experimental data in section \ref{results}. Our concluding remarks are given in
 section \ref{concl}.

\section{Experiment}\label{experiment}

Polycrystalline sample Pr$_{0.9}$Sr$_{0.1}$MnO$_3$ were prepared using Sol-Gel method from initial pure oxides; Pr$_6$O$_{11}$, SrCO$_{3}$ and Mn$_2$O$_3$, with nominal purities higher than 99.9\%. Identification of the crystalline phase and structural analysis were carried out by Powder X-ray diffraction (XRD) with a diffractometer (SIEMENS D500) using $\lambda_{CuKa}$ ($\lambda = 1.54056 \AA$) radiations at room temperature with a scan step of 0.02$^{\circ}$ in the range $20^{\circ} \leq 2\theta \leq 80^{\circ}$. Crystal structure parameters were refined using FullProf program  based on the Rietveld method [48,49]. Magnetization measurements versus temperature (T) and magnetic applied field used for MCE studies, were performed using the SQUID magnetometer developed in Louis Neel Laboratory at Grenoble, France.

Let us detail the sample preparation: polycrystalline  Pr$_{0.9}$Sr$_{0.1}$Mn$_{0.9}^{3+}$Mn$_{0.1}^{4+}$O$_{3}$ oxides were prepared by Sol–Gel method, and single perovskite phase was obtained by appropriate heat treatment in accordance with the following reaction:
0.15 Pr$_6$O$_{11}$ + 0.1 SrCO$_3$ + 0.5 Mn$_2$O$_3$  $\rightarrow$  Pr$_{0.9}$Sr$_{0.1}$MnO$_3$ +$\delta$ CO$_2$.

The appropriate amounts of high purity (99.9\%) Pr$_6$O$_{11}$, SrCO$_3$ and Mn$_2$O$_3$ powder precursors were dissolved in nitric acid and distilled water. The solution was then heated at 90°C on a hot plate under constant stirring to eliminate the excess water and to obtain a homogeneous solution. Subsequently, citric acid and ethylene glycol were added under thermal agitation to obtain a viscous gel. The gel was then dried at 300°C and calcinated at 600°C for 5 hours resulting in fine powder. Finally, the powders were ground and mixed in a mortar for 10 min and then placed in platinum crucibles and compressed into pellets 12 mm in diameter and 2 mm thick. The pellets were finally sintered at 1100°C for 3 hours.\\

Experimental results will be shown in section \ref{results} together with the MC results for comparison. 

\section {Model and Method}\label{model}
\subsection{Model}

\begin{figure}[ht]
\centering
\includegraphics[width=6cm,angle=0]{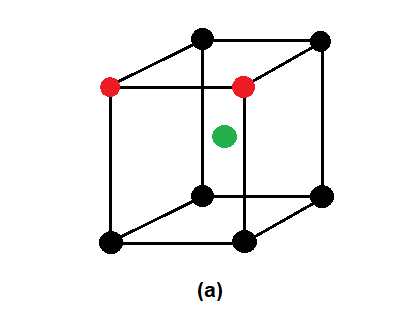}
\includegraphics[width=6cm,angle=0]{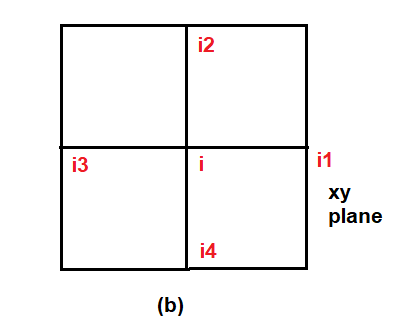}
\caption{(a) An example of ion distribution on the bct lattice: black, red and green circles represent Mn$^{3+}$, Mn$^{4+}$ and Pr$^{3+}$, respectively, (b) Spin of the Mn ion at the site $i$ interacts simultaneously with four Mn spins at sites $i1$, $i2$, $i3$ and $i4$ in the multi-spin interaction given by Eq. (\ref{multi}). See text for comments.  }
\label{fig1}
\end{figure}

\indent We consider the body-centered tetragonal (bct) lattice with the following pairwise-interactions (see Fig. \ref{fig1}a):
\begin{equation}\label{pairwise}
 {\cal H}_p = -\sum_{<i,j>}J_{ij}\mathbf S_{i}\cdot \mathbf S_{j}-\mu_0H\sum_{<i>}S_{i}
\end{equation}
where $\mathbf S_i$ is the spin at the lattice site $i$, $\sum_{<i,j>}$ is made over spin pairs coupled through
the exchange interaction $J_{ij}$.  $H$ is a magnetic field applied along the $z$ axis. 

Before defining explicitly the interactions, let us discuss about the spin model we shall use. We suppose that the spins of Mn ions lie on the $z$ axis with a strong uniaxial anisotropy.  We have first tried to calculate using the Heisenberg model with a strong anisotropy but we did not get an agreement with experimental data.  On the other hand, using a discrete Ising-like spin model, we obtain an excellent agreement with the experimental magnetization  in the whole temperature range as shown in the next section. So, we shall use the Ising spins with different  spin amplitudes depending on the ion kind.  We shall write $S_i$ instead of $\mathbf S_i$ etc in the following. Note that $S_i$ represents the Ising spin of amplitude $S$.

Let us recall that there are two kinds of Mn ions which occupy the corner sites of the bct lattice  in Pr$_{0.9}$Sr$_{0.1}$Mn$_{0.9}^{3+}$Mn$_{0.1}^{4+}$O$_{3}$: Mn$^{4+}$ with spin amplitude $S=3/2$ and Mn$^{3+}$ with $S=2$. Due to the doping, the positions of Mn$^{3+}$ and Mn$^{4+}$ are at random. The Pr and Sr ions occupy the centered sites of the bct lattice with their respective concentrations.   It is experimentally found that interaction between neighboring Mn$^{3+}$ and Mn$^{4+}$ as well as that between Mn$^{3+}$ and Mn$^{3+}$ are  strongly ferromagnetic while that between Mn$^{4+}$ and Mn$^{4+}$ is weakly antiferromagnetic. 
Hereafter we shall take the following interactions between nearest-neighbors (NN) between different kinds of magnetic Mn ions:

\indent $J_{1}$: Interaction coupling of a Mn$^{3+}$ ion with a NN Mn$^{3+}$ ion,\\ 
\indent $J_{2}$: Interaction coupling of a Mn$^{3+}$ ion with a NN Mn$^{4+}$ ion,\\  
\indent $J_{3}$: Interaction coupling of a Mn$^{4+}$ ion with a NN Mn$^{4+}$ ion,\\  
\indent $J_{4}$: Interaction coupling of a Pr ion with a Mn$^{3+}$ ion,\\
\indent $J_{5}$: Interaction coupling of a Pr ion with a Mn$^{4+}$ ion\\
\indent $J_{6}$: Interaction coupling between two Pr ions on the adjacent bct units.\\

Note that the two outer electrons of the Pr ion occupy the orbitals 4f2, so its spin is 1 (Hund's rule), and $L=3+2=5$. So, for less than half filled shell , $J=|L-S|=4$. However, as we are interested in spin-spin interactions of the compound which are responsible for the magnetic transition (not $J-J$ interaction), the orbitals $L$ are not taken into account.
 
It is known, by the theory of critical phenomena, that the nature of a phase transition depends on a few parameters:  the spin nature (Ising, XY, Heisenberg, Potts,..), the nature of their interaction and the space dimension. Only when the interaction is short-range ferromagnetic (or non frustrated antiferromagnetic), the transition obeys the universality class depending only on the nature of spin and the space dimension: one has 2D Ising universality class, 3D Ising universality class, 3D XY universality class, 3D Heisenberg universality class.... However, when a disorder is introduced, or a frustration due to competing interactions takes place ... the nature of the transition changes . It may have other critical exponents in the presence of a disorder, or it becomes of the first order  when  the system is frustrated  (see recent reviews in Ref. [50]). 
The transition experimentally observed for our compound is of the second order with a magnetization plateau below $T_c$. Our compound has a smalll disorder but there is no frustration. For the modeling, we have tried the Heisenberg model, the continuous Ising model, ... with various pairwise interactions, but we did not find the magnetization plateau and the sharp second-order transition experimentally observed. The idea to introduce the multispin interaction to reproduce the experimental observations, as we will see later, was successful.   After several trial forms, the following term is found to well describe the experimental magnetization curve:

\begin{equation}\label{multi}
 {\cal H}_m = - K\sum_{i} S_{i} S_{i1} S_{i2} S_{i3} S_{i4}
\end{equation}
where $K$ is the interaction strength and the sum runs over all Mn sites and the spins $ S_{i1}, \ S_{i2},\  S_{i3}$ and $ S_{i4}$ are the NN of the spins $S_i$ on the $xy$ plane (see Fig. \ref{fig1}b). 

Let us discuss about the multi-spin interaction given in Eq. (\ref{multi}).  First, we take just a plaquette shown in Fig. \ref{fig1}b: we see that  there are many degenerate configurations: all spins are parallel, any two spins among $i1$,$i2$, $i3$ $i4$  are reversed and  four spins $i1$,$i2$, $i3$ $i4$ are reversed. Of course, when the plaquette is connected to surrounding plaquettes and when we take into account the pairwise interactions,  the degeneracy is reduced. Nevertheless, this deneneracy favors a sharp second-order transition experimentally observed for the  compound under study.  At low temperatures, the multi-spin interaction favors the ferromagnetic state against  single-spin excitations. This explains the magnetization plateau experimentally observed.  

At this stage, it is worth to emphasize  that  the Heisenberg model is the lowest order involving only two spins because of the initial assumption of the overlap between the spin-dependent wave functions of only two neighboring atoms (Hartree-Fock approximation, see Ref. 51, pp. 55-60). In materials, however the interaction between one spin with its neighbors is simultaneous, but the demonstration for the multispin interaction as in the case of two-body Heisenberg model is at present impossible. There was an attempt to make a fourth-order pertubation expansion for the Heisenberg model. It results in a term of 3-spin  and a term of 4-spin interactions [52].  But for the Ising model, the demonstration by exact methods has been done. The reader is referred to Ref. [53] for the references on the demonstration of various multispin interactions. Note that recently there has been an increasing number of papers using the multispin interaction for various purposes [53-55].

 We have conducted standard MC simulations on  samples of dimension $N=L\times L\times L$, where $L$ is the number of bct cells in each of the $x$, $y$ and $z$ directions. Periodic boundary conditions are used in all directions.
 Simulations have been carried out for different lattice sizes ranging from $12^3$ to $30^3$ lattice cells to check finite-size effects.  The results shown below are those of $30^3$ lattice size (the finite-size effect is no more significant from $20^3$).

  The procedure of our simulation can be split into two steps. The first step consists in equilibrating the lattice at a given temperature.
The second step, when equilibrium is reached, we determine thermodynamic properties by taking thermal averages of various physical quantities [56,57]. Starting from a random spin configuration as the initial condition for the MC simulation, we have calculated
the internal energy per spin $ E$, the specific heat $C_V$, the magnetic susceptibility $\chi$, the magnetization of each sublattice and the total magnetization, as functions of temperature $T$ and magnetic field $H$.
The MC run time for equilibrating is about $10^5$ MC steps per spin. The averaging is taken, after equilibrating, over $10^5$ MC steps. A large number of runs have been carried out to check the reproductivity of results shown below.

 The statistiical averages of the $z$ spin component of Mn$^{3+}$ and Mn$^{4+}$ and Pr$^{3+}$ ($\langle S_1 \rangle$, $\langle S_2\rangle$ and $\langle S_3 \rangle$, respectively)  and the Edwards-Anderson order parameter $Q_{EA}$ for these three sublattices $\ell=1,2,3$ are defined by
\begin{eqnarray}
\langle S_{\ell} \rangle&=&\frac{1}{N_{\ell}}\langle\sum_{i\in \ell}S_i\rangle \label{msub}\\
Q_{EA}(\ell)&=&\frac{1}{N_{\ell}}\sum_{i\in \ell}\langle  S_i\rangle \label{qeas}
\end{eqnarray}
where $\langle...\rangle$ indicates the statistical time average and the sum is taken over Mn$^{3+}$ ($\ell=1$) or Mn$^{4+}$ ($\ell=2$) or Pr$^{3+}$ ($\ell=3$), with $N_{\ell}$ being the number of spins of each kind. Note that $Q_{EA}$ is calculated by first taking the time average of each spin and secondly taking the spatial average over all spins. This parameter is used to calculate the freezing degree of the spins when a long-range ordering is absent or the nature of ordering is unknown such as in spin glasses or in disordered systems [57-60].

The total magnetization $M$ is defined by
\begin{equation}
\langle M\rangle =g\mu_B (\langle S_{1}\rangle +\langle S_{2}\rangle+\langle S_{3}\rangle) \label{mtot}
\end{equation}
where $g$ and $\mu_B$ are the effective gyromagnetic factor and the Bohr magneton. Note that experimental data on the total magnetization (magnetic moment) is given by $M$. The magnetizations of Mn3+, Mn4+ and Pr3+ were not measured separately, therefore the gyromagnetic factor $g$ which relates the total spin to $M$ should be understood as an "effective $g$".

The average internal energy $E$ per spin, the specific heat $C_V$ per spin and the susceptibility $\chi$ per spin are defined by
\begin{eqnarray}
\langle E \rangle &=&\frac{1}{N}\langle ({\cal H}_p+{\cal H}_m)\rangle \\
C_V&=&\frac{1}{k_BT^2}[\langle E^2\rangle - \langle E\rangle ^2]\\
\chi&=&\frac{1}{k_BT} [\langle M^2\rangle -\langle M\rangle^2]
\end{eqnarray}


\section{Results - Comparison with Experiment } \label{results}

\subsection{How to fit the MC results with experiments}

Experiments found that $J_1$ and $J_2$  dominate and give rise to the ferromagnetic ordering up to very high temperatures $T_C=220$ K for  Pr$_{0.9}$Sr$_{0.1}$Mn$_{0.9}^{3+}$Mn$_{0.1}^{4+}$O$_{3}$. For MC simulations, after many trial sets of parameteers, we found the  following set which  reproduces the experimental magnetization: 

\begin{eqnarray}
J_1&=&+6.0J,\  J_2= +6.0J,\  J_3=-0.1J,\  J_4= -6.0J, \ J_5= -3.0J, \ J_6= -0.1J \label{val1}\\
C&=&0.5  \label{val2}\\
K&=&1.9  \label{val3}
\end{eqnarray}
where $J$ is equal  to 1, namely $J$ is the MC energy unit, $C$ is the reduction coefficient applied to the interaction between Pr-Pr on the $z$ axis. 

\begin{figure}[ht]
\centering
\includegraphics[width=8cm,angle=0]{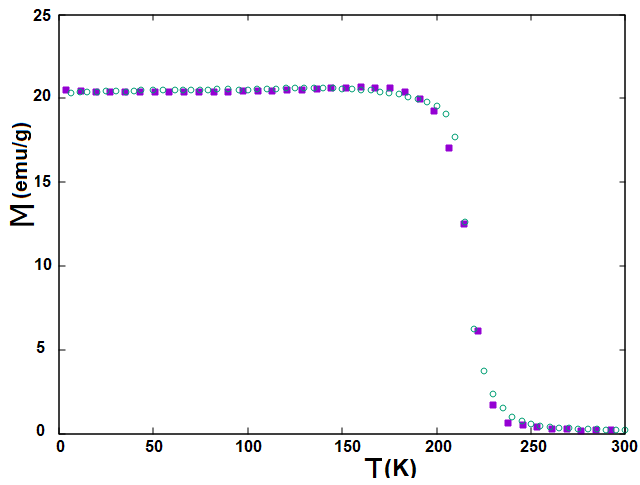}
\caption{MC result (violet squares) and experimental magnetization (green void circles) versus  temperature $T$ in Kelvin, are shown for comparison. See text for comments.  }
\label{mag}
\end{figure}

 Using the above interaction values we obtained the MC transition temperature $T_C(MC)=57.11$.  
In order to fit the MC transition temperature with the experimental value $T_C(exp)=220$ K, we have to multiply all the interaction values given above by $J=220/57.11$. Let us show the MC curve and the experimental magnetization in Fig. \ref{mag}.

We note the following important points:

(i) The experimental magnetization shows a flatness up to the phase transition temperature. This is unusual in magnetic materials. At first, we have tried to modify the values of $J_1,..., J_6$ to get the agreement, but we failed. At best the MC curve coincides with the experimental one up to $T=200$ K and starts to decrease from that temperature to make the transition at $T=220$K.  The deficit of the MC magnetization between 200 and 220 K with respect to the experimental one is solved by introducing the multispin interaction between Mn ions on the $xy$ plane: this finally gives the good MC result in excellent agreement with the experimental magnetization. 

(ii) The fall of the magnetization curve at $T_C$ is very sharp. However, as there  is no discontinuity of the magnetization at $T_C$, the transition is thus of second-order. This is confirmed by the internal energy and the specific heat which will be shown below.

 In order to estimate the amplitudes of  physical parameters in real units, we 
 use the mean-field approximation [50]:
\begin{equation}\label{MFTC}
T_C=\frac{2}{3k_B}ZS_{eff}(S_{eff}+1)J_{eff}
\end{equation}
where $Z=6$Mn+$8\times 0.9$Pr=13.72  is effective coordination number at a Mn  site  and $S_{eff}$  the effective spin length which can be taken as the average on the Mn$^{4+}$, Mn$^{3+}$ and Pr$^{3+}$ using their concentrations: $S_{eff}=(0.9\times2+0.1\times1.5+0.9\times 1)/(0.9+0.1+0.9)=1.5$. Putting $T_C=220$ K in Eq. (\ref{MFTC}), we obtain $J_{eff}\simeq =0.0005525$ eV=5.52 K. 

From the fit of $T_C$ above, it is easy to determine each of the exchange interactions defined earlier, in Kelvin. For example,  $J_1=6J=6\times220/57.11\simeq 23$ K.   Note that in magnetic materials with Curie temperatures at room or higher temperatures the exchange interaction is of the order of several dozens of Kelvin [61,62] which is the same order of magnitude  as what we found here.  Other interactions can be calculated in the same manner. Note that using such a mean-field approximation, we obtain  the order of magnitude of interaction parameters, but not to a good precision. 

Now, what is the real value of the energy in the real unit ? To answer this question, there are two ways to do:

1. We use the classical ground-state energy with the effective parameters calculated above:

\begin{eqnarray}
E_0(MF)&=&-0.5\times Z\times S_{eff}^2 \times  J_{eff}\\
&=&-0.5\times 13.72\times1.5^2 \times 55.25\times10^{-5}\  \mbox{eV}\\
&=&-852.73\times10^{-5} \mbox{ eV}\\
&\simeq&-8.53\  \mbox{meV}\label{MFE}
\end{eqnarray}

2. We use the  MC result of the GS energy using the interaction values given in Eqs. (\ref{val1})-(\ref{val3}) which is $E_0/J\simeq 68$ when $T\rightarrow 0$. This value has been obtained with $k_B=1$ to simplify the MC simulation. To recover the real value of $E_0$, we write

\begin{eqnarray}
E_0(MC)&=&-68\times 3.8582\times k_B\\
&=& -262.3576\times 1.380649\times 10^{-23} \ \mbox{Joules}\\
&=&-362.21\times10^{-23}\times 6.242\times 10^{18}\  \mbox{eV}\\
&=&-2260.92\times10^{-5}\  \mbox{ eV}\\
&\simeq&-22.61\  \mbox{meV}\label{MCE}
\end{eqnarray}
This value is almost three times higher than that of the mean-field approximation in Eq. (\ref{MFE}). It is interesting to calculate the effective exchange interaction from this value: one has

\begin{eqnarray}
               J_{eff}&=&  E_0 (MC)/[-0.5\times Z_{eff}\times  S_{eff}^2 ]\\
                      & =&22.61/[-0.5\times13.72\times1.5^2] \ \mbox{meV}\\
                      &=& 1.4648 \ \mbox{meV}\\
                      &\simeq& 14.46 \ \mbox{K} 
\end{eqnarray}
This value is of course more reliable than the value 5.52 K deduced from the mean-field approximation.  It is in the range of exchange interactions found in magnetic materials [60] and in another family of perovskite compound [61].

\begin{figure}[ht]
\centering
\includegraphics[width=6cm,angle=0]{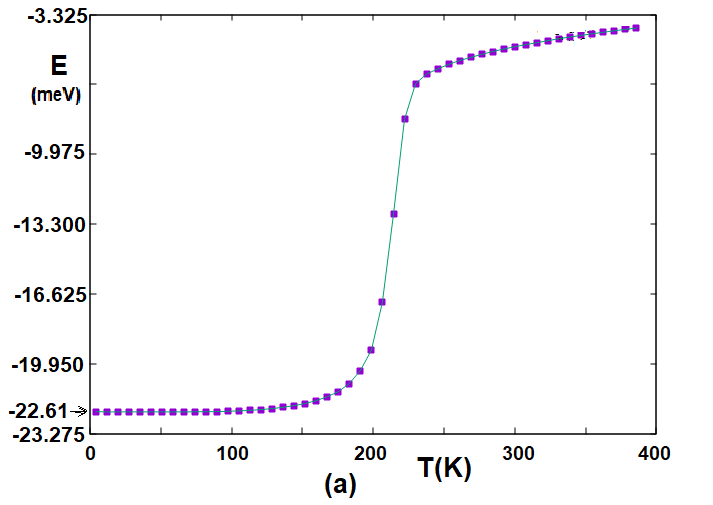}
\includegraphics[width=6cm,angle=0]{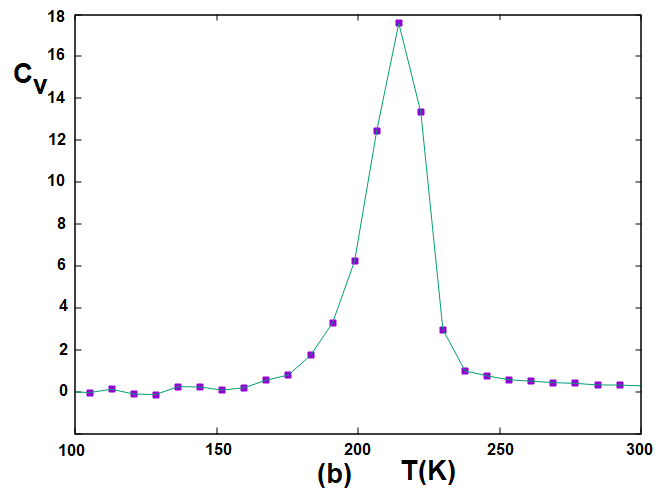}
\caption{(a) Internal energy per spin (in meV) versus  temperature $T$ in Kelvin, (b) Specific heat  (in meV/K) versus $T$. Line is guide to the eye.}
\label{fig3}
\end{figure}

 Let us show the  MC energy $E$  versus  temperature $T$(K) and the specific heat $C_V$ versus $T$  in Fig. \ref{fig3}. 

We note that

(i) The $E$ curve shows a change of curvature at $T=220$ K which confirms the value of $T_C$ in Fig. \ref{mag}. There is no sign of discontinuity of $E$ at $T_C$ confirming that the transition is of second order. Note that in a first-order transition, the order parameter (magnetization) and the internal energy are discontinuous at $T_C$ [ 63,64]. This is not the case here.

(ii) The energy at $T=0$ is -22.61 meV.

(iii) The peak of $C_V$ confirms the change of curvature of $E$ at $T_C$.  The rather large width of $C_V$ at $T_C$ and the finite peak height confirm again the second-order nature of the transition.

\begin{figure}[ht]
\centering
\includegraphics[width=6cm,angle=0]{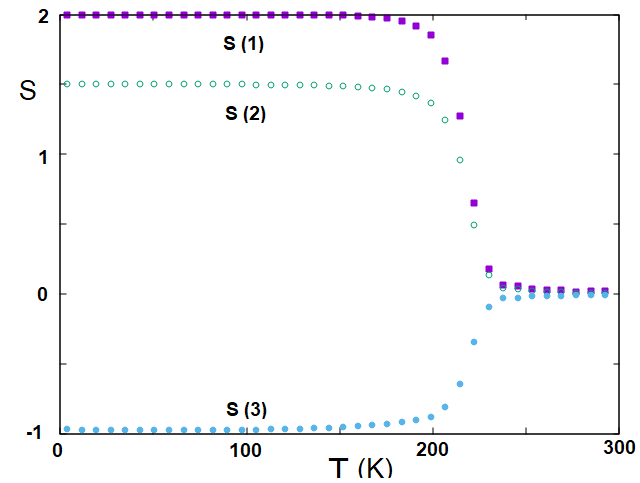}
\caption{ Sublattice $z$ spin components $S(1)$, $S(2)$ and $S(3)$   versus  temperature $T$.  }
\label{fig4}
\end{figure}

We show now in Fig. \ref{fig4} the $z$ component three sublattice magnetic ions $\langle S (i) \rangle\ (i=1,2,3)$.   We see here that the Mn$^{3+}$ ($S(1)$) and  Mn$^{4+}$ ($S(2)$) have the same sign, namely they order ferromagnetically, while $S(3)$ (Pr ions) is ordered antiferromagnetically with the Mn ions.  These data are MC results, they were not available by experiments.

\begin{figure}[ht]
\centering
\includegraphics[width=6.6cm,angle=0]{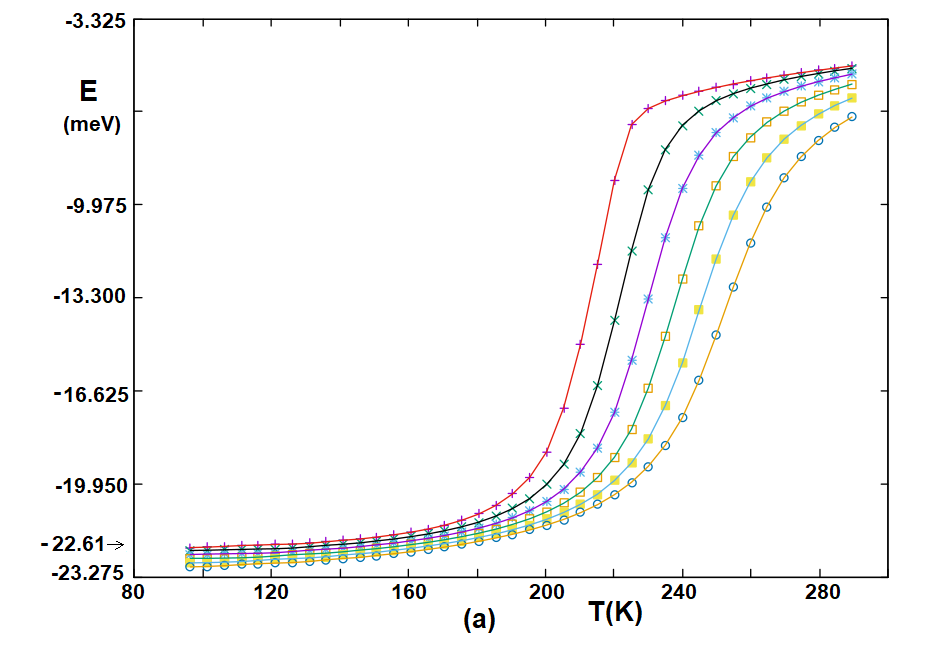}
\caption{ Energy versus $T$  for $\mu_0H=0.03, 1, 2, 3, 4, 5$ Tesla (from left to right). See text for comment }
\label{fig5}
\end{figure}

We show in Fig. \ref{fig5} the MC result of the internal energy for magnetic field $\mu_0H$ ranging from 0.03 to 5 Tesla.

 As seen in Fig. \ref{fig5}, the transition temperature increases with increasing $H$, and the transition is less and less sharp.  This is well known in magnetic systems under an applied field.  Rigorously speaking, there is no  phase transition in a ferromagnet under an applied magnetic field since the global magnetization never goes to zero at finite $T$.

\subsection   {Magnetocaloric Effect - Magnetic Entropy Change}

\begin{figure}[ht]
\centering
\includegraphics[width=11cm,angle=0]{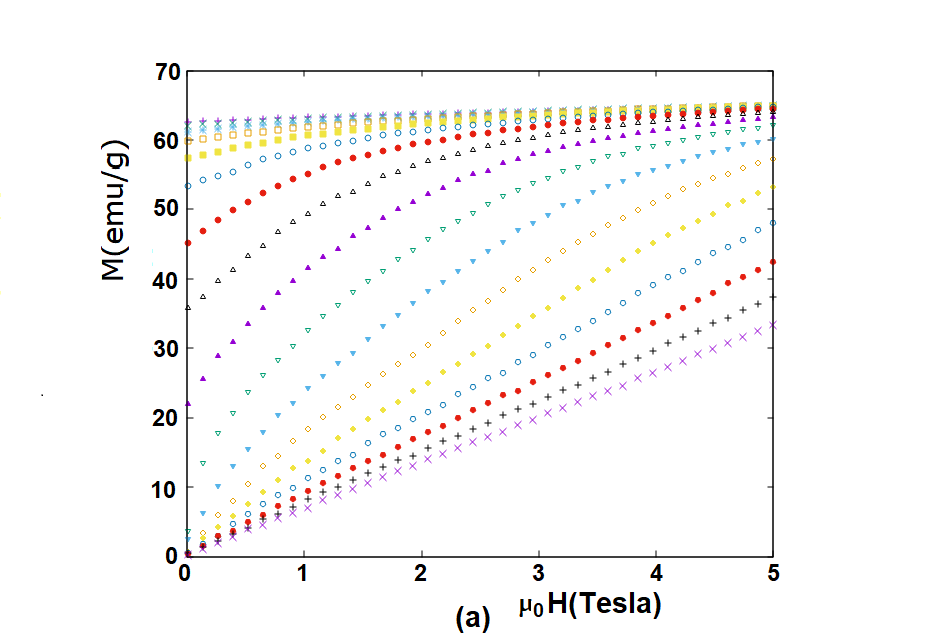}
\includegraphics[width=13cm,angle=0]{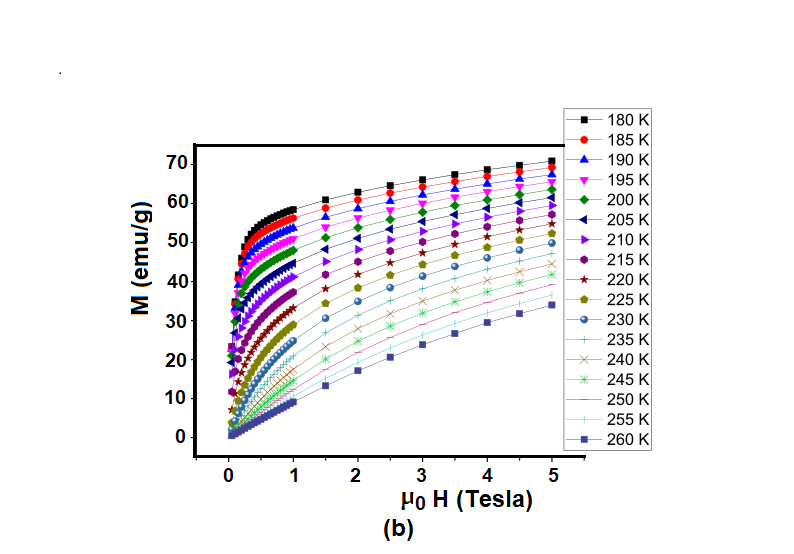}
\caption{ Effect of magnetic field $\mu_0 H$ (Tesla) on the magnetization (a) obtained from MC simulations, from top line : $T=180$ K, 185 K, 190 K, ..., 260 K ( bottom line), (b) Experimental data for the same system in the same temperature range indicated on the figure.  }
\label{fig6}
\end{figure}

We apply the magnetic field on the system at a given $T$. The curves are shown in Fig. \ref{fig6}a for $T$ below and above $T_C$. For clarity, experimental data in this range of $T$ are separately shown in Fig. \ref{fig6}b for comparison. We obtain a good qualitative agreement between the two sets of curves.  We have some remarks:

i) The MC magnetization curves far below $T_C$ are larger than the  experimental curves at low $H$. We think that this is because experimental samples are polycrystalline ones which have domains resulting in low $M$ at low $H$ and low $T$, in contrast to  MC samples which are on a lattice,

ii) For $T$ close to $T_C$ and above $T_C$, the agreement between experiments and MC results is quite good.  

\begin{figure}[ht]
\centering.
\includegraphics[width=7cm,angle=0]{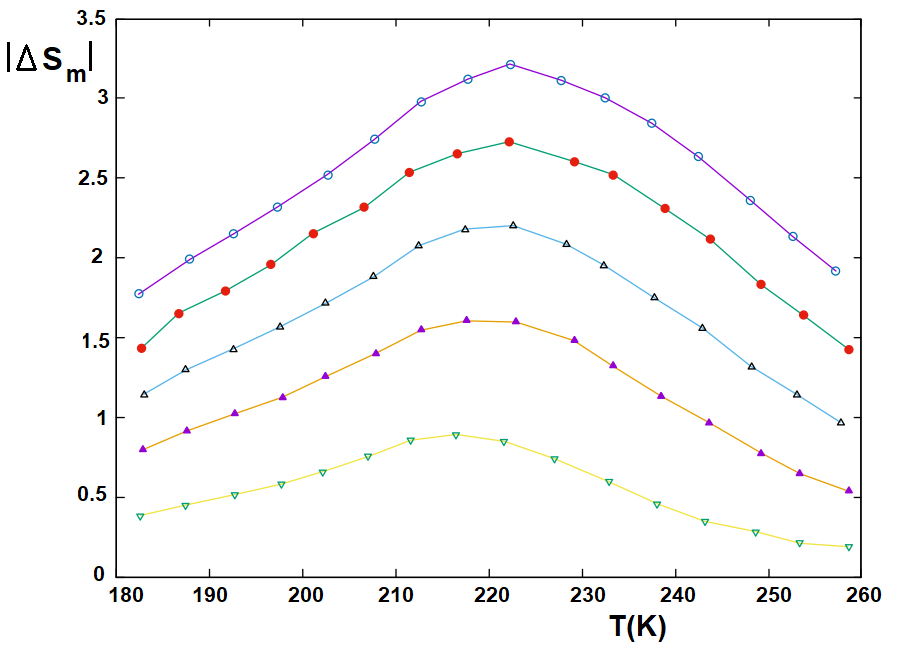}
\caption{ Experimental magnetic entropy change $|\Delta S_m|$ (J/(kgKelvin) versus  temperature $T$ for $\mu_0H=$1, 2, 3, 4, 5 Tesla (from bottom to top). See text for comment.}
\label{figDeltaS}
\end{figure}

Let us show now the results  on the magnetocaloric effect of our compound.  This effect is characterized by  the magnetic entropy change, called $|\Delta S_m|$, and the Relative Cooling Power (RCP).   
Let us note that $|\Delta S_m|$ is not the statistical entropy defined for a given $(T,H)$.  $|\Delta S_m|$, the magnetic entropy change, is  calculated by applying a field progressively from 0 to $ \mu_0 H$ at a given temperature. $|\Delta S_m|$ is given by the  thermodynamic Maxwell formula

\begin{equation}\label{DeltaS}
|\Delta S_m (T,H)|=\int_0^H\left[\frac{\delta M(T,H_i)}{\delta T} \right]_{H_i}\  \mu_0 dH_i
\end{equation}
where $\delta M$ is the change of the magnetization at $H$ when $T$ varies $T\rightarrow T+\delta T$.  This formula is discretized  and used in experiments as

\begin{equation}\label{DeltaS1}
|\Delta S_m (T,H)|=\sum_i\left[\frac{M_i-M_{i+1}}{T_{i+1}-T_i} \right]\mu_0 \Delta H_i
\end{equation}
The experimental magnetic entropy change $|\Delta S_m|$ using this formula   is shown in Fig. \ref{figDeltaS}: the peak temperature increases very slightly with increasing $H$ but the peak height is higher with larger $H$. The large peaks  of $|\Delta S_m|$ indicate the smootth change from the ferromagnetic phase to paramagnetic phase. This is a common feature of doped perovskite compounds which have more or less domains, defects and structural inhomogeneities [65].

\begin{figure}[ht]
\centering.
\includegraphics[width=7cm,angle=0]{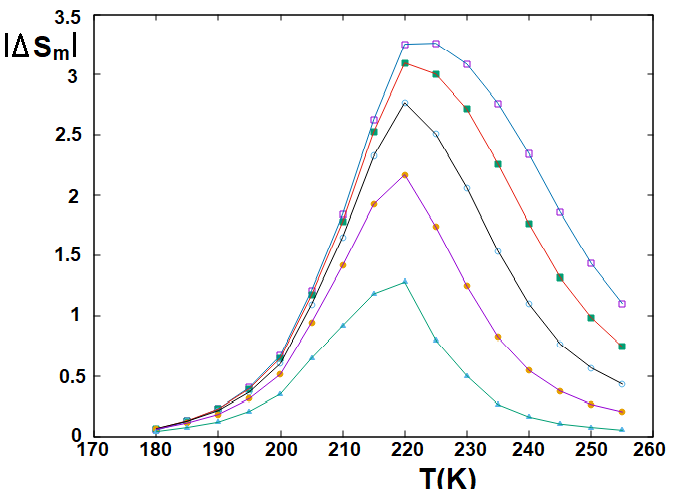}
\caption{ MC result for magnetic entropy change $|\Delta S_m|$ (J/(kgKelvin) versus  temperature $T$ for $\mu_0H=$1, 2, 3, 4, 5 Tesla (from bottom to top). See text for comment.}
\label{figMCDeltaS}
\end{figure}

To compare with experimental magnetic entropy change, we carry out the calculation of  $|\Delta S_m|$ using Eq. (\ref{DeltaS1}) as in experiments with the MC magnetization shown in Fig. \ref{fig6}a.  This gives the result shown in Fig. \ref{figMCDeltaS}.  Several remarks are in order:

i) $|\Delta S_m|$ obtained from MC simulations show a peak at each value of $\mu_0H$ from 1 Tesla to 5 Tesla. The value of the peak increases with increasing $H$, in agreement with experiments shown in Fig. \ref{figDeltaS}. The peak temperarure slightly increases with increasing $H$ as observed in experiments.

ii) Unlike experiments, the MC $|\Delta S_m|$  comes to zero below $T\simeq 180$ K. This is due to the fact that the MC magnetization is higher than the experimental one at low $T$ and low $H$ as said earlier. Below $T=180$ K the MC system is highly magnetized so that low $H$ does not make effect.  Unlike experimental samples, the MC system is on a lattice,  there are thus no vacancies, no dislocations ... The system is, though disordered,  more homogeneous so that the peak widths are narrower.
 
The formula used to calculate the RCP is 

\begin{equation}\label{RCP}
RCP(H)=|\Delta S_{max} (H)|\times \Delta T
\end{equation}
where $|\Delta S_{max} (H)|$ is the maximum value of $|\Delta S_{m} (H)|$ and  $\Delta T$  the temperature range at  the full
width at half maximum.  

We show in Table \ref{figRCP} our experimental Relative Cooling Power (RCP) and the MC RCP. As seen, there is a difference between experiments and MC results. This difference comes from very large temperature range at the full width at half maximum in experimental data (Fig. \ref{figDeltaS}) as discussed above.  Nevertheless, our compound studied here, both experimentally and numerically,  has very high RCP which is very important for high efficiency in applications using the magnetocaloric effect.


\begin{table}[ht]
\centering
\vspace{-1cm}
\includegraphics[width=10cm,angle=0]{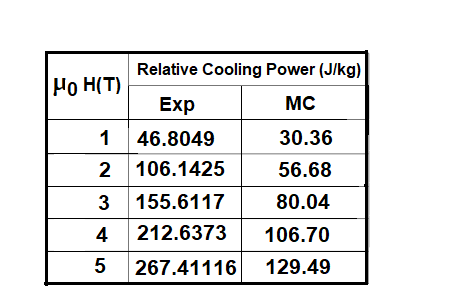}
\vspace{1cm}
\caption{ Experimental Relative Cooling Power and MC results of Pr$_{0.9}$Sr$_{0.1}$Mn$_{0.9}^{3+}$Mn$_{0.1}^{4+}$O$_{3}$, for  $\mu_0H=1,2,3,4,5$ Tesla. See text for comments. }
\label{figRCP}
\end{table}

\section{Conclusion}\label{concl}

We have shown in this paper magnetic properties of the perovskite compound Pr$_{0.9}$Sr$_{0.1}$Mn$_{0.9}^{3+}$Mn$_{0.1}^{4+}$O$_{3}$ observed experimentally. These include the magnetization as functions of $T$ and $H$ and the magnetic entropy change at the transition from the ferromagnetic phase to the disordered phase at $T_C\simeq 220$ K.  We deduce the RCP and find that it varies with $H$ to a very high value, very interesting for magnetocaloric applications.   To model this compound to fit with the experimental magnetization which shows a plateau up to the transition temperature, we introduce a five-spin interaction term between the Mn ions in the Hamiltonian, in addition to the different pairwise interactions between magnetic ions Mn$^{3+}$, Mn$^{4+}$ and Pr$^{3+}$. MC simulations have been carried out. We have found an excellent agreement between the MC magnetization and the  experimental one. We have deduced from the fitting the values of various exchange interactions. Other physical quantities such as internal energy and specific heat  have been calcultated as functions of $T$. 

We have also compared the experimental magnetic entropy change $|\Delta S_{m} (H)|$ with the results from MC simulations. There is an excellent agreement on the peak values of $|\Delta S_{m} (H)|$. However, for a given $H$ the experimental result show a broad  maximum while the MC one show a sharper peak. As a consequence, the MC RCP is lower than the experimental one.  

The overall agreement found in this paper between experiments and the theoretical model elaborated for the compound  Pr$_{0.9}$Sr$_{0.1}$Mn$_{0.9}^{3+}$Mn$_{0.1}^{4+}$O$_{3}$ is remarkable. The multi-spin interaction introduced here is new, it allows to reproduce the experimental magnetization plateau below $T_C$ and other experimental data.   Work is underway to study the case of other concentrations of Pr-Sr and of Mn${^{3+}}$-Mn${^{4+}}$.

\vspace{2cm}

\acknowledgments

Yethreb Essouda  is indebted to the CY Cergy Paris University for hospitality during her working visits.

\end{document}